# Simulation of interstitial diffusion of ion-implanted boron

O.I. Velichko and N.V. Kniazhava

**Abstract**

A model of the interstitial diffusion of ion-implanted boron during rapid thermal annealing of silicon layers previously amorphized by implantation of germanium has been proposed. It is supposed that the boron interstitials are generated continuously during annealing due to dissolution or rearrangement of the clusters of impurity atoms, which are formed in the ion-implanted layers with impurity concentration above the solubility limit. The local elastic stresses arising due to the difference of boron atomic radius and atomic radius of silicon also contribute to the generation of boron interstitials. On the basis of the model proposed a simulation of redistribution of ion-implanted boron during rapid thermal annealing with duration of 60 s at a temperature of 850 degrees Celsius has been carried out. The calculated profile of boron distribution after thermal treatment agrees well with the experimental data that confirms the adequacy of the model. A number of the parameters of interstitial diffusion have been derived. In particular, the average migration length of nonequilibrium boron interstitials is equal to 12 nanometers. It was also obtained that approximately 1.94 % of boron atoms were converted to the interstitial sites, participated in the fast interstitial migration, and then became immobile again transferring into a substitutional position or forming the electrically inactive complexes with defects of crystal lattice.

**PACS:** 61.72.Cc; 61.72.Ji; 61.72.Tt; 66.30.Jt; 07.05.Tp

*Keywords:* Boron; silicon; ion implantation; annealing; diffusion

**Введение**

Для создания современных полупроводниковых приборов и интегральных микросхем широко используются методы легирования посредством ионной имплантации в сочетании с последующей термической обработкой полупроводниковых слоев. Схематическое изображение этого способа легирования приведено на Рис.1.

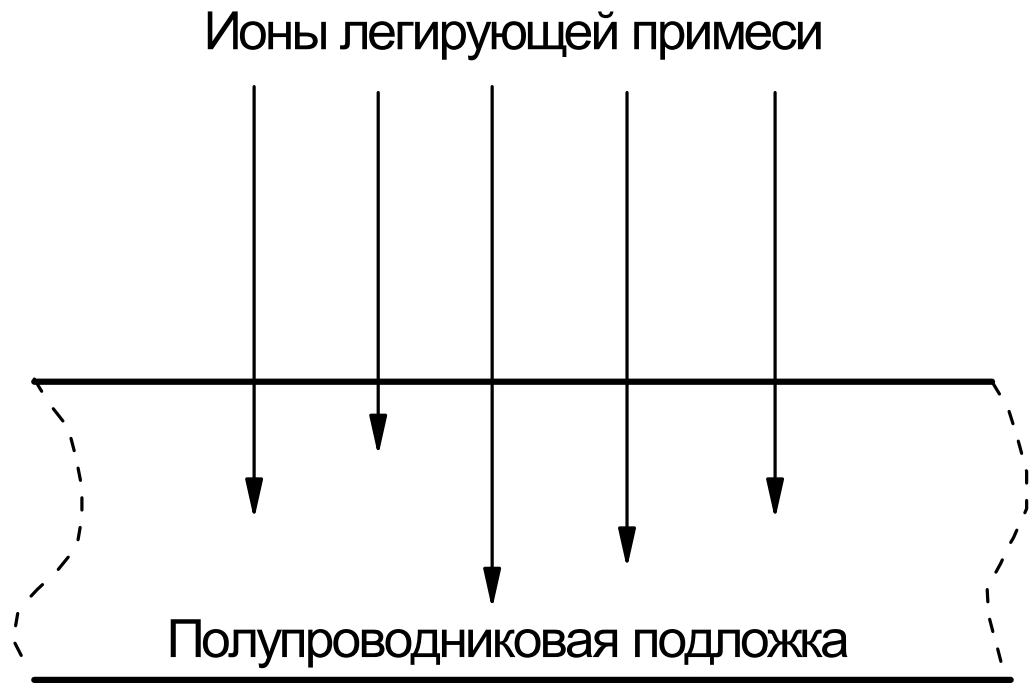

Рис.1. Легирование посредством ионной имплантации

В процессе ионного внедрения происходит генерация большого количества неравновесных постимплантационных дефектов, в том числе вакансий, собственных межузельных атомов, а также межузельных атомов примеси. Для удаления постимплантационных дефектов и электрической активации примеси применяется термическая обработка при температурах ~700 $^o$C и выше. В процессе термообработки происходит диффузионное перераспределение примесных атомов. Для уменьшения этого перераспределения и улучшения электрофизических параметров приборов широко применяются быстрые [1-3] или даже пиковые (spike annealing) высокотемпературные отжиги [1,4], либо короткие [5,6] или быстрые [7] низкотемпературные обработки. Как следует из экспериментальных данных, в этих случаях может наблюдаться длиннопробежная миграция неравновесных межузельных атомов примеси, что проявляется в формировании характерных протяженных "хвостов" в области низкой концентрации примесных атомов [1,5-7]. Типичные профили распределения концентрации атомов ионно-имплантированного бора после низкотемпературного термического отжига, полученные в работах [5] и [7], представлены на Рис.2 и Рис.3 соответственно. Необходимо отметить, что профиль распределения атомов бора, представленный на Рис.3, создавался ионной имплантацией с энергией ионов 500 эВ с целью обеспечить формирование слоев нанометровых размеров, пригодных для изготовления интегральных микросхем с ультравысокой степенью интеграции (Ultra Large Scale Integration — ULSI).

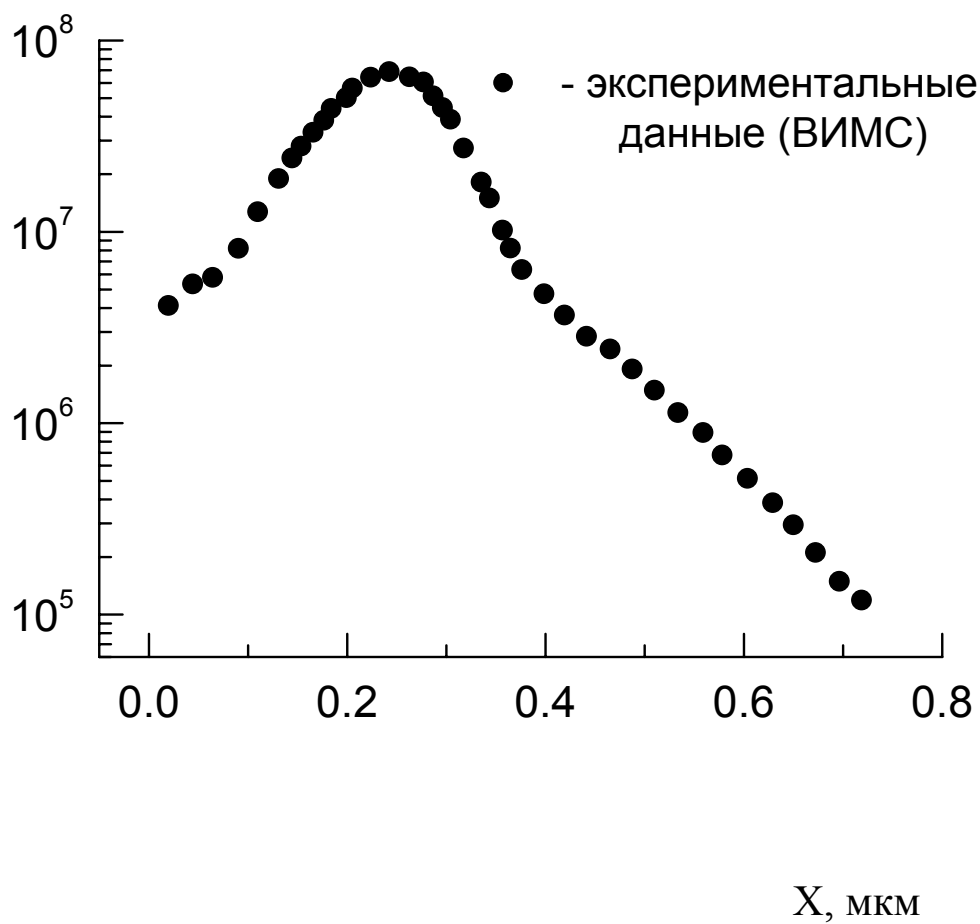

Рис.2. Профиль распределения концентрации атомов ионно-имплантированного бора после термического отжига в течение 35 минут при температуре 800 °C. Энергия имплантации 70 кэВ, доза $10^{15}$ см$^{-2}$.

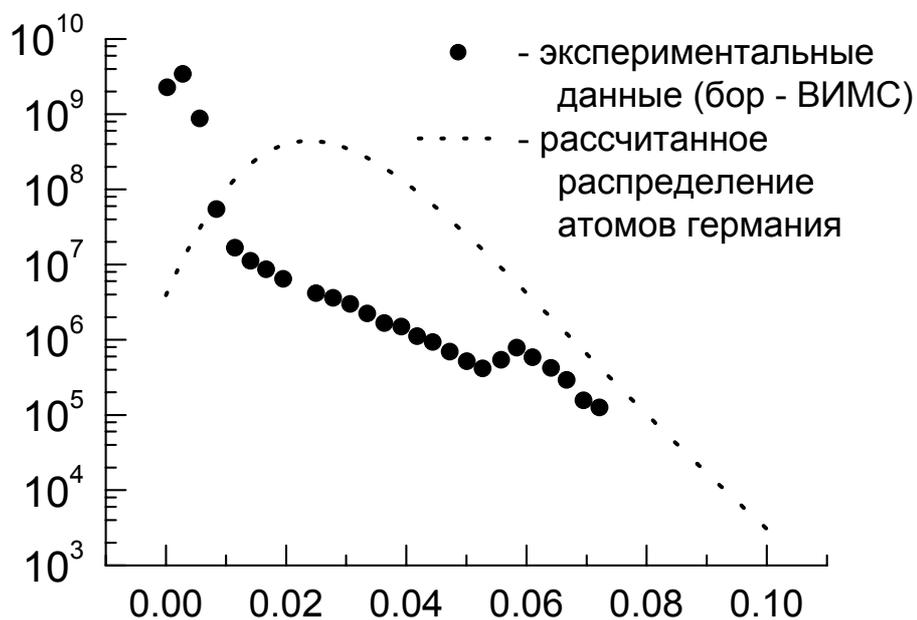

Рис.3. Профиль распределения концентрации атомов ионно-имплантированного бора после термического отжига кремниевых подложек с предварительно аморфизованным слоем в течение 60 секунд при температуре 850 °C. Энергия имплантации 500 эВ, доза $2\times10^{15}$ см$^{-2}$. Пунктирной линией изображен рассчитанный профиль распределения концентрации германия после имплантации с энергией 32 кэВ и дозой $1\times10^{15}$ см$^{-2}$.

В работе [5] для экспериментов использовались образцы кристаллического кремния с удельным сопротивлением 16500 Ом см, которые имплантировались ионами бора. Энергия имплантации 70 кэВ, доза $10^{15}$ см$^{-2}$. Отжиг проводился в атмосфере бора при температуре 800 $^{\circ}$C в течение 35 минут.

В работе [7] для экспериментов использовались подложки кремния *n*-типа ориентации <100> с удельным сопротивлением 10-25 Ом см, выращенные по методу Чохральского. Подложки имплантировались ионами германия Ge с энергией 32 кэВ и дозой $1\times10^{15}$ см$^{-2}$, что имеет следствием создание аморфного слоя глубиной 55 нанометров. Глубина аморфного слоя определялась методом Резерфордовского обратного рассеяния и просвечивающей электронной микроскопией поперечного сечения (XTEM). После внедрения Ge подложки имплантировались ионами бора с энергией 500 эВ и дозой $2\times10^{15}$ см$^{-2}$. После имплантации бора проводился изохронный отжиг подложек при 850 $^{\circ}$C длительностью 60 с. Распределение общей концентрации бора определялось методом вторичной ионной масс-спектроскопии (ВИМС). Анализ электрической активации образцов после отжига осуществлялся путем измерения слоевого сопротивления и дозы активной примеси посредством эффекта Холла. Наличие и структура протяженных дефектов определялись просвечивающей электронной микроскопией поперечного сечения с использованием слабого пучка.

**Как видно из Рис.2 и Рис.3, даже при таких низких температурах отжига, имеет место существенная диффузия атомов бора в области с концентрацией меньшей $5\times10^{6}$ мкм$^{-3}$ или $2\times10^{7}$ мкм$^{-3}$ соответственно.**

**Существующие коммерческие программы моделирования процессов легирования полупроводников (см. например, [8,9]) не описывают этого явления.**

В работе [10] было предположено, что образование протяженных "хвостов" в области низкой концентрации примесных атомов, которое имеет место при перераспределении ионно-имплантированной примеси в случае коротких низкотемпературных термообработок, есть следствие длиннопробежной миграции неравновесных межузельных атомов примеси. Такие неравновесные межузельные атомы примеси могут образовываться в ионно-имплантированных слоях в результате того, что часть примесных атомов оказывается в межузельном положении непосредственно после имплантации ионов, либо переходит в межузельное положение вследствие распада части имплантационных дефектов на начальной стадии отжига. Графическая интерпретация процесса такой длиннопробежной миграции неравновесных межузельных атомов примеси представлена на Рис.4.

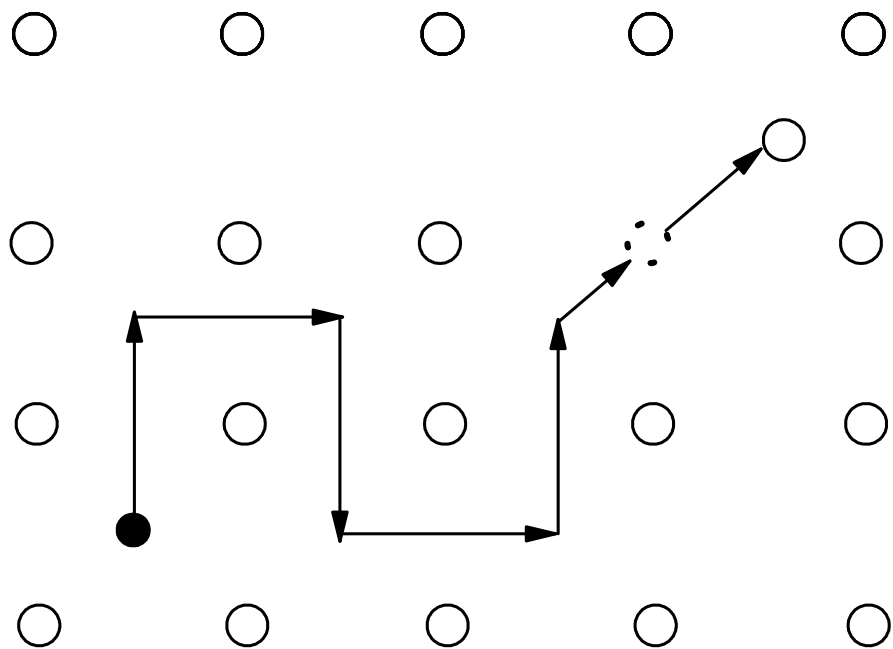

Рис. 4. Схематичное представление механизма прямой межузельной миграции атомов примеси.

Моделирование процесса перераспределения ионно-имплантированного бора, представленного на Рис.2, выполненное в работе Величко и Бурунова [11] в рамках предположения о длиннопробежной миграции неравновесных межузельных атомов примеси, показало прекрасное соответствие рассчитанного профиля распределения концентрации примеси экспериментальным данным. Используем поэтому данное предположение для расчета процесса перераспределения ионно-имплантированного бора при коротком низкотемпературном отжиге, представленного на Рис.3.

# ЦЕЛЬ ИССЛЕДОВАНИЯ

**Проведенный анализ позволяет сформулировать цель данного исследования:**

**1. Основываясь на предположении о длиннопробежной миграции неравновесных межузельных атомов примеси, разработать модель межузельной диффузии ионно-имплантированного бора в слоях кремния предварительно аморфизованных имплантацией ионов германия при низкотемпературных отжигах секундной длительности.**

**2. Провести моделирование процесса перераспределения ионно-имплантированного бора при низкотемпературном отжиге секундной длительности.**

**3. Определить характерные значения параметров межузельной диффузии ионно-имплантированного бора.**

## 1. Анализ предыдущих исследований и постановка задачи

В работе Величко и Бурунова [11] было проведено моделирование процесса перераспределения ионно-имплантированного бора при температуре отжига 800 $^{o}$C длительностью 35 минут. Доза имплантации составляла $10^{15}$ см$^{-2}$. Поскольку ионы бора являются достаточно легкими (бор занимает 5 место в периодической таблице Менделеева), использованная доза имплантации является недостаточной для создания аморфного слоя. Это означает, что имплантированный слой содержит достаточно большое количество постимплантационных дефектов, включая межузельные атомы бора. Кроме того, межузельные атомы бора генерируются на начальной стадии отжига в результате распада постимплантационных дефектов, содержащих бор. В работе [11] рассматривалась миграция этих неравновесных межузельных атомов бора в течение всей термообработки. С этой целью было получено решение нестационарного уравнения диффузии межузельных атомов примеси.

В отличие от процесса, рассмотренного в [11], в работе [7] имплантация ионов бора осуществляется в слой, предварительно аморфизованный более тяжелыми ионами германия (германий занимает 32 место в периодической таблице Менделеева). Это означает, что на начальной стадии отжига происходит твердофазная рекристаллизация аморфного слоя, в результате которой удаляются дефекты в пределах аморфного слоя, а значительная часть атомов бора занимает положение замещения, даже если их концентрация превышает предел растворимости. Как правило, рекристаллизованный слой является весьма совершенным с точки зрения отсутствия постимплантационных дефектов.

Небольшая часть оставшихся постимплантационных дефектов расположена в объеме полупроводника за пределами границы области рекристаллизованного аморфного слоя, то есть на глубине примерно 0.055 мкм и более. Результаты проведенного в [7] анализа дефектной структуры рекристаллизованных подложек с помощью просвечивающей электронной микроскопии подтверждают это предположение.

Как показывают экспериментальные данные [12], процесс низкотемпературной твердофазной рекристаллизации предварительно созданного аморфного слоя, обеспечивая отсутствие постимплантационных дефектов в пределах области рекристаллизации, не приводит тем не менее к 100 % активации примеси при высоких дозах имплантации бора. Так, в работе Landi et al. [12] исследовался процесс активации ионно-имплантированного бора в результате твердофазной рекристаллизации аморфных слоев, созданных высокодозной имплантацией кремния. Было обнаружено, что в результате твердофазной рекристаллизации при температуре 600 $^o$C в случае концентрации примеси выше $3.5 \times 10^8$ мкм$^{-3}$ происходит образование электрически неактивного бора. Это значение концентрации существенно выше предела растворимости бора $C_{sol}$, который для рассматриваемой температуры отжига 600 $^o$C составляет $5.63 \times 10^6$ мкм$^{-3}$ [13]. Таким образом, хотя значение концентрации электрически активного бора существенно выше предела растворимости, часть атомов примеси является электрически неактивной, образуя кластеры атомов бора с дефектами кристаллической структуры, например с межузельными атомами кремния [14-16]. Следует ожидать, что описанное явление будет иметь место и для рассматриваемой температуры отжига 850 $^o$C. Предел растворимости бора $C_{sol}$ для данной температуры равен $4.88 \times 10^7$ мкм$^{-3}$ [13]. Как видно из Рис.3, это значение значительно (почти на 2 порядка) меньше максимальной концентрации примесных атомов после имплантации ионов $C_m \approx 4.44 \times 10^9$ мкм$^{-3}$, то есть для рассматриваемого процесса отжига значительная часть атомов бора после твердофазной рекристаллизации является электрически активной, находясь к тому же выше предела растворимости, а другая значительная часть бора электрически неактивна, будучи связанной в кластеры. Как показывают экспериментальные данные, при температурах отжига выше 750 $^o$C имеет место процесс деактивации атомов бора, то есть атомы примеси, концентрация которых превышает предел растворимости, также начинают связываться в кластеры. Таким образом, для рассматриваемого процесса отжига при 850 $^o$C после твердофазной рекристаллизации имеет место образование новых кластеров атомов бора. Кроме того, может происходить перестройка уже имеющихся кластеров, например в результате процесса созревания по Оствальду (Ostwald ripening) [17]. Суть этого процесса заключается в исчезновении кластеров малого размера и увеличении количества кластеров более крупных размеров. Таким образом, в процессе отжига после твердофазной рекристаллизации происходит образование новых кластеров атомов бора и перестройка уже имеющихся кластеров бора в результате распада кластеров малого размера и роста кластеров более крупных размеров. В результате такого распада или перестройки часть примесных атомов может перейти в межузельное положение [18].

Необходимо также учесть, что внедрение бора проводится в слой, аморфизованный имплантацией германия. В процессе твердофазной рекристаллизации аморфного слоя имплантированные атомы германия занимают положение в узлах кристаллической решетки вместо части атомов кремния. В результате в рекристаллизованном слое возникают упругие напряжения, поскольку атомный радиус германия $r_A^{Ge}$ =0.139 нм (ковалентный радиус 0.122 нм) больше атомного радиуса кремния $r_A^{Si}$ =0.134 нм (ковалентный радиус 0.117 нм) [19]. На Рис.3 пунктирной линией изображено распределение концентрации атомов германия, рассчитанное с использованием распределения Пирсон-IV [20]. В соответствии с таблицами [21] для энергии имплантации германия 32 кэВ и дозы $1\times10^{15}$ см$^{-2}$ были использованы следующие значения параметров, задающих распределение имплантированных атомов: $C_m^{Ge}$ = 4.1556×10$^8$ мкм$^{-3}$; $R_p$ = 0.0258 мкм; $\Delta R_p$ = 0.0096 мкм; $Sk^{Ge}$ = 0.486.

Здесь $C_m^{Ge}$ — максимальная концентрация атомов германия;

$R_p^{Ge}$ — средний проективный пробег ионов германия;

$\Delta R_p^{Ge}$ — страгглинг проективного пробега ионов германия;

$Sk^{Ge}$ — асимметрия распределения атомов германия.

С другой стороны, из Рис.3 видно, что в приповерхностном слое глубиной 0.005 мкм возникают упругие напряжения противоположного знака, поскольку атомный радиус бора $r_A^B$ =0.095 нм (ковалентный радиус 0.089 нм) меньше атомного радиуса кремния $r_A^{Si}$ =0.134 нм (ковалентный радиус 0.117 нм) [19]. Эти напряжения возникают из-за увеличения объема между атомами кристаллической решетки в области имплантации ионов бора вследствие малости атомного радиуса бора. Согласно работе [22] это явление имеет следствием генерацию межузельных атомов.

## 2. Модель процесса межузельной диффузии

### 2.1. Качественная модель межузельной диффузии

Основываясь на результатах предыдущего анализа предполагаем, что в области, в которой концентрация атомов ионно-имплантированного бора превышает предел растворимости, происходит генерация неравновесных межузельных атомов бора. Генерация этих неравновесных частиц происходит в результате образования, перестройки или распада кластеров атомов бора, а также следствие действия напряжений, обусловленных малостью атомного радиуса бора по сравнению с радиусом атома кремния. В отличие от работы Величко и Бурунова [11], предполагается, что процесс генерации межузельных атомов происходит не только на начальной стадии, а непрерывно в течение всего быстрого отжига (в нашем случае это 60 с). Длиннопробежная миграция неравновесных межузельных атомов бора из области высокой концентрации ионно-имплантированной примеси обеспечивает формирование протяженного "хвоста" в области низкой концентрации. Для описания процесса межузельной диффузии и расчета концентрационного профиля распределения атомов бора после имплантации и быстрого термического отжига можно использовать следующую систему дифференциальных уравнений.

### 2.2 Исходные уравнения

Поскольку, согласно нашим предположениям, генерация неравновесных межузельных атомов бора происходит непрерывно, используем для описания генерации, диффузии и рекомбинации межузельных атомов бора систему уравнений, полученную в [23] для описания миграции неравновесных межузельных атомов водорода, непрерывно генерируемых в приповерхностной области полупроводника при обработке кремниевых подложек в водородосодержащей плазме газового разряда. Применительно к рассматриваемому процессу диффузии бора эта система имеет следующий вид:

**1. Закон сохранения неподвижных атомов примеси**

$$\frac{\partial C^T(x,t)}{\partial t} = \frac{C^{AI}(x,t)}{\tau^{AI}} - G^{AI}(x,t), \tag{1}$$

**2. Уравнение диффузии неравновесных межузельных атомов бора**

$$d^{AI}\frac{\partial^2 C^{AI}}{\partial x^2} - \frac{C^{AI}(x,t)}{\tau^{AI}} + G^{AI}(x,t) = 0, \qquad (2)$$

или

$$\frac{\partial^2 C^{AI}}{\partial x^2} - \frac{C^{AI}}{l_{AI}^2} + \frac{\widetilde{g}^{AI}(x,t)}{l_{AI}^2} = 0, \qquad (3)$$

где

$$l_{AI} = \sqrt{d^{AI}\tau^{AI}}, \qquad \widetilde{g}^{AI}(x,t) = G^{AI}(x,t)\tau^{AI}. \qquad (4)$$

Здесь $C^T$ — суммарная концентрация атомов примеси в положении замещения и связанных в кластеры (неподвижные атомы примеси) $C^{AI}$ — концентрация неравновесных межузельных атомов примеси; $d^{AI}$ и $\tau^{AI}$ — коэффициент диффузии и среднее время жизни неравновесных межузельных атомов примеси, соответственно; $G^{AI}$ — скорость генерации межузельных атомов примеси.

Мы используем стационарное уравнение диффузии для межузельных атомов примеси ввиду их большой средней длины миграции ($l_i >> l_{fall}$, где $l_{fall}$ — характерная длина заметного уменьшения концентрации примеси) и из-за небольшого время жизни этих неравновесных межузельных атомов ($\tau_{AI} << \tau_p$, где $\tau_p$ — продолжительность термической обработки).

Система уравнений (1) и (3) описывает диффузию примеси в результате миграции неравновесных межузельных атомов примеси. Чтобы решить эту систему уравнений, нужны соответствующие граничные условия. Рассмотрим одномерную область (1D) конечной длины $[0, x_B]$ и добавим к уравнению (3) граничные условия непротекания на поверхности полупроводника

$$d^{AI} \left.\frac{\partial C^{AI}}{\partial x}\right|_{x=0} = 0 . \tag{5}$$

и граничные условия первого рода в объеме полупроводника

$$\left. C^{AI} \right|_{x=x_B} = C_B^{AI} . \tag{6}$$

Добавим также начальные условия

$$C^T(x,0) = C_0(x) , \qquad C^{AI}(x,0) = C_{eq}^{AI} = const \tag{7}$$

к уравнению (1) и уравнению (3) соответственно, где

$C_0(x)$ — распределение неподвижных атомов бора после твердофазной рекристаллизации (предполагается, что это распределение совпадает с распределением имплантированной примеси);

$C_{eq}^{AI}$ — равновесное значение междоузельной концентрации атомов примеси (предполагается, что $C_{eq}^{AI}$ равно нулю для рассматриваемого случая диффузии).

Чтобы описать пространственное распределение атомов примеси после твердофазной рекристаллизации и пространственное распределение скорости генерации межузельных атомов бора, используем распределение Гаусса

$$C_0(x) = C_m \exp\left[-\frac{(x-R_p)^2}{2\Delta R_p^2}\right], \tag{8}$$

$$G^{AI}(x,t) = g_m \exp\left[-\frac{(x-R_p)^2}{2\Delta R_p^2}\right], \tag{9}$$

где

$$C_m = \frac{Q}{\sqrt{2\pi}\Delta R_p} \times 10^{-8} \ . \tag{10}$$

Здесь $C_m$ — максимальное значение концентрации атомов бора после имплантации; $g_m$ — максимальное значение скорости генерации межузельных атомов бора в единице объема полупроводника; $Q$ — доза имплантированных атомов примеси ион/см$^{-2}$; $R_p$ и $\Delta R_p$ — средний проективный пробег ионов бора и страгглинг проективного пробега, соответственно.

### 2.3. Используемый метод расчета распределения примеси

Используем для расчета распределения атомов ионно-имплантированного бора после термического отжига аналитическое решение краевой задачи (1), (3), (5)-(9), полученное в работе Величко и Соболевская [23]. Применительно к рассматриваемому случаю перераспределения ионно-имплантированного бора это решение имеет вид

$$C^T(x,t) = \frac{1}{\tau^{AI}} \int_0^t C^{AI}(x,t)dt + (1 - p^{AI})C_0(x) \ , \tag{11}$$

$$\begin{aligned}C^{AI}(x,t) = C_m^{AI} \frac{\exp u_1}{\cosh u_2^B} \{\cosh u_2 \, [\exp(-u_6) \\ \times (erfu_4^B - erfu_4) + \exp(u_6)(erfu_5^B - erfu_5)] \\ + \exp(-u_9)\sinh u_3 \times [erfu_7 + \exp(2u_9)(erfu_5 - erfu_8) - erfu_4]\} \\ + C_B^{AI} \frac{\cosh u_2}{\cosh u_2^B}, \end{aligned} \tag{12}$$

где

$$C_m^{AI} = \frac{\sqrt{\pi}\, g_m \tau^{AI} \Delta R_p}{2\sqrt{2}\, l_{AI}} \ , \tag{13}$$

$$u_1 = \frac{\Delta R_p^2}{2l_{AI}^2} , \qquad (14)$$

$$u_2 = \frac{x}{l_{AI}} , \qquad u_2^B = \frac{x_B}{l_{AI}} , \qquad (15)$$

$$u_3 = \frac{x_B - x}{l_{AI}} , \qquad (16)$$

$$u_4 = \frac{\Delta R_p^2 - l_{AI} R_p + l_{AI} x}{\sqrt{2}\Delta R_p l_{AI}} , \qquad u_4^B = \frac{\Delta R_p^2 - l_{AI} R_p + l_{AI} x_B}{\sqrt{2}\Delta R_p l_{AI}} , \qquad (17)$$

$$u_5 = \frac{\Delta R_p^2 + l_{AI} R_p - l_{AI} x}{\sqrt{2}\Delta R_p l_{AI}} , \qquad u_5^B = \frac{\Delta R_p^2 + l_{AI} R_p - l_{AI} x_B}{\sqrt{2}\Delta R_p l_{AI}} \qquad (18)$$

$$u_6 = \frac{R_p - x_B}{l_{AI}} , \qquad (19)$$

$$u_7 = \frac{\Delta R_p^2 - l_{AI} R_p}{\sqrt{2}\Delta R_p l_{AI}} , \qquad u_8 = \frac{\Delta R_p^2 + l_{AI} R_p}{\sqrt{2}\Delta R_p l_{AI}} , \qquad (20)$$

$$u_9 = \frac{R_p}{l_{AI}} . \qquad (21)$$

Здесь $p^{AI}$ — доля атомов примеси, переходивших в межузельное положение.

Для расчета рассматриваемого процесса легирования кремния бором была использована разработанная ранее на языке высокого уровня Fortran программа моделирования процессов межузельной диффузии примесных атомов при термообработках легированных слоев.

## 3. Моделирование процесса легирования

На базе разработанной модели с использованием аналитического решения краевой задачи межузельной диффузии, полученного в [23], было проведено моделирование процесса перераспределения атомов ионно-имплантированного бора в кремнии в результате межузельной диффузии неравновесных атомов примеси при быстром термическом отжиге. Для сравнения использованы экспериментальные данные [7]. Как указывалось ранее, длительность обработки 60 с, температура отжига — 850 $^O$C. Результаты расчета концентрационного профиля распределения бора представлены на Рис.5.

Как видно из Рис.5, профиль распределения бора, рассчитанный с использованием выражений (11) и (12), находится в хорошем согласии с экспериментальными данными [7]. При расчете использовались следующие значения параметров разработанной модели межузельной диффузии ионно-имплантированной примеси, которые обеспечивают наилучшее соответствие рассчитанной кривой экспериментальному профилю концентрации бора после термообработки.

**Параметры, задающие начальное распределения атомов бора:** $R_p$=0.0026 мкм (2.6 нм); $\Delta R_p$=0.0018 мкм (1.8 нм).

**Параметры, описывающие процесс межузельной диффузии:** средняя длина пробега межузельных атомов бора $l^{AI}$=0.012 мкм (12 нм); значение скорости генерации неравновесных межузельных атомов примеси в максимуме распределения $g_m$=1.56×10$^6$ мкм$^{-3}$с$^{-1}$.

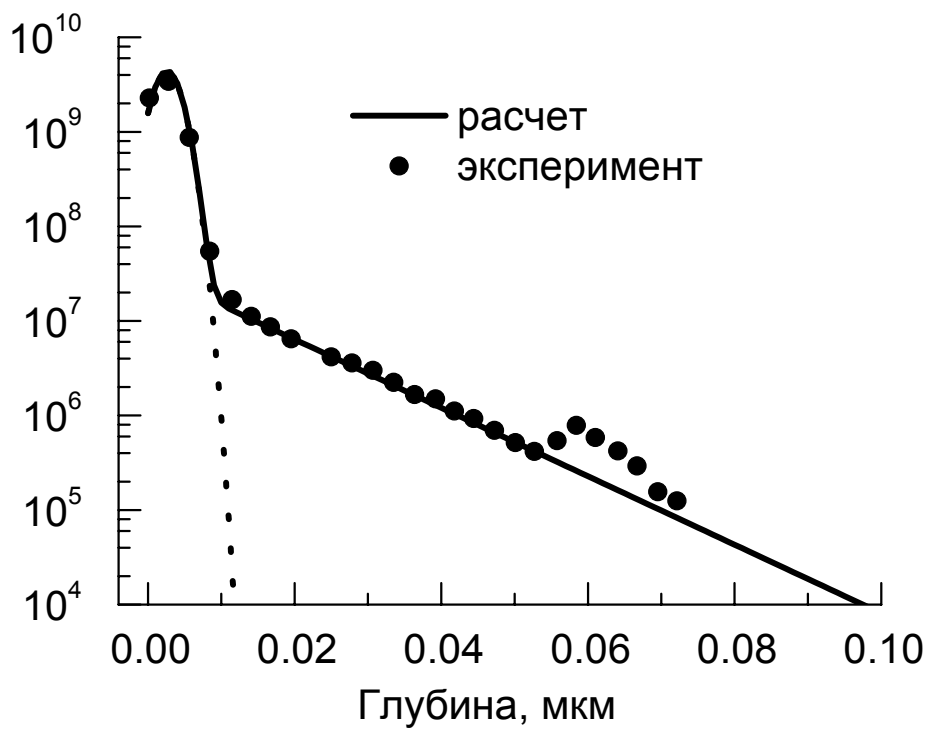

Рис. 5. Рассчитанное распределение атомов бора в кремнии, сформированное в результате межузельной диффузии неравновесных атомов примеси. Длительность обработки 60 с, температура отжига — 850 $^O$C. Пунктирная кривая — рассчитанное распределение бора после имплантации. Экспериментальные данные (•) взяты из работы Hamilton et al. [7].

Как следует из проведенных расчетов, приблизительно 1.94 % атомов бора переходит в межузельное положение, мигрируют и затем опять становятся неподвижными, переходя в положение замещения или образуя электрически неактивные комплексы с дефектами. Миграция этих неравновесных межузельных атомов примеси во время проведения термического отжига приводит к формированию "хвоста" на профиле распределения бора в области падения концентрации примеси. Как видно из представленных результатов моделирования, расчет процесса перераспределения ионно-имплантированного бора, выполненный с использованием разработанной модели межузельной диффузии ионно-имплантированной примеси, хорошо согласуется с экспериментальными данными. Отклонение результатов расчета от экспериментальных данных на глубинах больших 0.055 мкм и более связано с поглощением межузельных атомов бора постимплантационными дефектами, оставшимися после твердофазной рекристаллизации. Учесть это явление предполагается в будущем исследовании.

# ЗАКЛЮЧЕНИЕ

**1.** Разработана модель межузельной диффузии ионно-имплантированного бора в кремнии при быстрых термических отжигах секундной длительности предварительно аморфизованных слоев. Предполагается, что межузельные атомы бора непрерывно образуются в течение отжига вследствие образования, перестройки или распада кластеров атомов примеси в ионно-имплантированных слоях с концентрацией примеси выше предела растворимости, а также в результате действия упругих напряжений, возникающих вследствие малости атомного радиуса бора по сравнению с атомным радиусом кремния.

**2.** Для решения сформулированной краевой задачи диффузии атомов бора было использовано полученное ранее аналитическое решение закона сохранения атомов примеси и уравнения межузельной диффузии.

**3.** Проведено моделирование процесса перераспределения ионно-имплантированного бора при быстром термическом отжиге длительностью 60 секунд при температуре 850 $^O$C. Рассчитанный профиль распределения атомов бора после термообработки хорошо согласуется с экспериментальными данными, что подтверждает адекватность разработанной модели.

**4.** Совпадение с экспериментальными данными позволило определить ряд параметров межузельной диффузии, в частности среднюю длину пробега неравновесных межузельных атомов бора, равную 0.012 мкм (12 нм) при температуре 850 $^O$C. Также установлено, что приблизительно 1.94 % атомов бора переходят в межузельное положение, участвуют в быстрой миграции и затем опять становятся неподвижными, переходя в положение замещения или образуя электрически неактивные комплексы с дефектами.

# СПИСОК ИСПОЛЬЗОВАННЫХ ИСТОЧНИКОВ